\documentclass[a4paper,twocolumn,
english,aps,prb,floatfix,groupedaddress,showpacs,nofootinbib]{revtex4-1}
\usepackage[T1]{fontenc}
\usepackage[latin1]{inputenc}
\usepackage{amsmath}
\usepackage{babel}
\usepackage{graphics}
\usepackage{amssymb}
\usepackage{mathrsfs}
\usepackage{dcolumn}


\begin{document}
\title
{Localization and spin transport in 
honeycomb structures with spin-orbit coupling}
\author {S. L. A. \surname{de Queiroz}}
\email{sldq@if.ufrj.br}
\affiliation{Instituto de F\'\i sica, Universidade Federal do
Rio de Janeiro, Caixa Postal 68528, 21941-972
Rio de Janeiro RJ, Brazil}

\date{\today}

\begin{abstract} 
Transfer-matrix methods are used for a tight-binding description of 
electron transport in graphene-like geometries, in the presence of 
spin-orbit couplings.  Application of finite-size scaling and 
phenomenological renormalization techniques shows that, for strong 
enough spin-orbit interactions and increasing on-site disorder, this 
system undergoes a metal-insulator transition characterized by the 
exponents $\nu=2.71(8)$, $\eta=0.174(2)$.  We show how one can extract 
information regarding spin polarization decay with distance from an 
injection edge, from the evolution of wave-function amplitudes in the 
transfer-matrix approach. For (relatively weak) spin-orbit coupling 
intensity $\mu$, we obtain that the characteristic length $\Lambda_s$ 
for spin-polarization decay behaves as $\Lambda_s \propto \mu^{-2}$.
\end{abstract}
\maketitle
 
\section{Introduction} 
\label{intro} 
In this paper we consider electronic transport in two-dimensional (2D) 
honeycombe-lattice structures with spin-orbit  (SO) interactions. 
While single-parameter scaling predicts that noninteracting electron states
in 2D are generally localized by disorder~\cite{gof4},
it is known that in this marginal dimension the enhancement of forward scattering 
provided by SO
effects can partially offset disorder-induced quantum interference, thus giving
rise to a conducting phase~\cite{hln80,b84,ando89}. Motivation for considering 
a honeycomb geometry is provided mostly by recent progress in experimental
synthesis, and theoretical understanding, of graphene~\cite{RMP,peres,ml10}.
It should be noted that some of carbon's cousin elements in 
group IVB have been predicted and, to varying degrees of success, shown 
to crystallize in a similar honeycomb arrangement, giving rise respectively 
to silicene~\cite{sil12} and germanene~\cite{ger15}. Stable honeycomb 
structures have also been found  for boron nitride 
(h-BN)~\cite{nov05} as well as SiO$_2$ (silicatene)~\cite{sand14} and
other elements or compounds. The ideas and methods exhibited in what follows
are, in principle, applicable to any of the above. However, where 
comparison of our results to experimental data is pertinent we generally 
restrict ourselves to graphene, as this is so far the best-understood
member of the ensemble.

Our purpose here is twofold: (i) to investigate location and universality
properties of the second-order metal-insulator found in these systems, 
for strong SO coupling, with increasing on-site disorder; and (ii) to examine
the spatial evolution of the polarization of an initially fully spin-polarized current, 
injected into one such  hexagonal-lattice system with SO couplings.

Section~\ref{sec:th} below recalls selected existing results, as well
as some technical aspects of the transfer-matrix (TM) method used in our
calculations. In Sec.~\ref{sec:mi} we discuss the metal-insulator transition. 
In Sec.~\ref{sec:relax} we show how  one can  extract information on the evolving state 
of polarization of an injected electron  beam, from the analysis of wavefunction
amplitudes generated  via the TM method.  
In Sec.~\ref{sec:conc} we provide a global analysis of our results; 
finally, concluding remarks are made.

\section{Theory}
\label{sec:th}

The model one-electron Hamiltonian for this problem can be written as
\begin{equation}
{\cal H}= \sum_{i,\sigma} \varepsilon_i\,c^\dagger_{i\sigma}\,c_{i\sigma} +
\sum_{\langle i,j\rangle}\,\sum_{\sigma, \sigma^\prime} V_{ij\,\sigma\sigma^\prime}
\,c^\dagger_{i\sigma}\,c_{j\sigma^\prime}\ ,
\label{eq:hdef}
\end{equation}
where $c^\dagger_{i\sigma}$, $c_{i\sigma}$ are creation and annihilation operators
for a particle with spin eigenvalues $\sigma=\pm1$ at site $i$, and the 
self-energies 
$\varepsilon_i$ are, in general, independently-distributed random variables;
$V_{ij\,\sigma\sigma^\prime}$ denotes the $2 \times 2$ spin-dependent hopping 
matrix between pairs of nearest-neighbor sites $\langle i,j\rangle$, whose 
elements must be consistent with the symplectic symmetry of SO
interactions.

Although modeling SO coupling in real systems usually relies on
including e.g. $p$ orbitals, with their corresponding degeneracies, 
in Eq.~(\ref{eq:hdef})
we resort to the customary description via an effective Hamiltonian with a 
single ($s-$like) orbital per site; for a pedagogically clear explanation 
of how this approach works from first principles, see e.g. the Appendix of 
Ref.~\onlinecite{ando89}.
Since we shall not attempt detailed numerical comparisons to experimental 
data, such formulation seems adequate for our purposes.

In the effective-Hamiltonian context, there is some leeway as the
specific form of the hopping term is concerned, depending on whether
one is specifically considering Rashba- or Dresselhaus- like couplings~\cite{ando89,kka08},
or the focus is simply  on the basic properties of systems in the
symplectic universality class~\cite{ez87,e95,aso04}. 
One constraint is that it must incorporate the basic symmetries of SO
coupling, which have close connections with quaternion algebra~\cite{mc92,mjh98}.

Here we use the implementation of Refs.~\onlinecite{ez87,e95}, namely:
\begin{equation}
V_{ij}=I+\mu i\sum_{k=x,y,z}V^k\sigma^k=
\begin{pmatrix}{1+i\mu V^z}&{\mu V^y+i\mu V^x}\cr
{-\mu V^y+i\mu V^x}&{1-i\mu V^z}\end{pmatrix}\ , 
\label{eq:vdef}
\end{equation}
where $I$ is the $2\times 2$ identity matrix, $\sigma^k$ are the Pauli
matrices, and $\mu$ gives the intensity of the SO coupling;
below we consider the (real) $\{V^k\}$ as either uniform ($V^x=V^y=V^z$ on
all bonds) or randomly (uniformly) distributed in $[-1/2,1/2]$. Thus
all energies will be written in units of the $\mu \equiv 0$ nearest-neighbor
hopping. 

Similarly the $\varepsilon_i$ will either be constant, site-independent,
or taken from a random uniform distribution in $[-W/2,W/2]$. 

The form Eq.~(\ref{eq:vdef}) for the hopping term does not exhibit the 
explicit multiplicative coupling between momentum and spin degrees of freedom,
characteristic of Rashba-like Hamiltonians~\cite{ando89,kka08}.  A similarly-decoupled
effective Hamiltonian can be derived from first principles for carbon nanotubes,
see Eqs. ~(3.15),~(3.16) of Ref.~\onlinecite{ando00}.
In two dimensions one should not expect significant discrepancies between 
results from either type of approach, as long as one is treating systems without 
lateral confinement.  

We apply the TM approach specific to tight-binding
Hamiltonians like Eq.~(\ref{eq:hdef})~\cite{ps81,ose68,ranmat} to 
finite-width strips of the honeycomb lattice, with $N$ sites across. 
Adaptation from the more usual square-lattice geometry is
straightforward, closely following the lines used for the 
TM description of localized (e.g., Ising and Potts) spin 
systems~\cite{pf84,bww90,dq06,dq13}. Two distinct orientations 
are possible~\cite{dq13,yy07}, with the TM proceeding either 
(a) perpendicularly~\cite{pf84} or (b) parallel to one lattice 
direction~\cite{bww90}. Case (a) corresponds to a
"brick" lattice, i.e., a square lattice with vertical bonds alternately 
missing. 

In the terminology of quasi one-dimensional carbon nanotubes (CNT)
and nanoribbons (CNR)~\cite{RMP}, a narrow strip with periodic boundary 
conditions across
in geometry (a) would be topologically equivalent to an armchair CNT, while
one in geometry (b) would correspond to a zigzag CNT. Conversely, free boundary
conditions parallel to the TM's direction of advance give: zigzag CNR in (a),  armchair CNR in (b).  
  
Detailed consideration shows that implementation of the 
TM scheme of Ref.~\onlinecite{ps81} in geometry (b) involves a number of
cumbersome intermediate operations [$\,$mostly matrix inversions, see
Eqs.~(8)--(16) of Ref.~\onlinecite{yy07}$\,$]. In what follows we
always make use of geometry (a), for simplicity.

We briefly recall selected aspects of the TM formulation introduced in 
Ref.~\onlinecite{ps81}, and of its adaptation for a honeycomb geometry with
SO couplings.

Consider a strip of the square lattice, cut along one of the coordinate directions.
For the orthogonal universality class
with site disorder, denoting by $k= 1, \dots, M$ the successive
columns, and $i = 1, \dots N$ the respective positions of
sites within each column of a strip, the recursion relation 
for an electronic wave function at  energy $E$ is given in terms 
of its local amplitudes (which can all be assumed real),
$\{\,a_{ik}(E)\,\}$, and tight-binding orbitals $|ik\rangle$,~as:
\begin{equation}
\begin{pmatrix}{\psi_{k+1}}\cr{\psi_k}\end{pmatrix}=
\begin{pmatrix}{P_k}&{-I}\cr{I}&{0}\end{pmatrix}
\begin{pmatrix}{\psi_k}\cr{\psi_{k-1}}\end{pmatrix}\  ,\ 
\psi_k \equiv
\begin{pmatrix}{a_{1k}}\cr {a_{2k}}\cr {\cdots}\cr       
{a_{Nk}}\end{pmatrix}\ \,
\label{eq:matrix}
\end{equation}
where $I$ is the $N \times N$ identity matrix, the energy dependence has been omitted for clarity,  and  (invoking periodic boundary conditions across the strip),
\begin{equation}
 {P_k} =\begin{pmatrix}{E-\varepsilon_{1k}}&{-1}&{\ \, 0}&{ \cdots}&{
-1}\cr
{-1 }&{ E-\varepsilon_{2k}}&{ -1}&{\cdots }&{\ 0}\cr
{\cdots }&{ \cdots }&{ }&{ }&{ -1}\cr
{-1 }&{\ 0 }&{ \cdots  }&{-1 }&{\ E-\varepsilon_{Nk}}\end{pmatrix}
\label{eq:pmatrix}\ .
\end{equation}
For a honeycomb lattice in the "brick" geometry,  the changes to $P_k$
are~\cite{pf84, bww90,dq06,dq13}: (i)  the off-diagonal elements are of the
form $-\left(1+(-1)^{i+k})\right)/2$, reflecting the alternately missing vertical bonds; 
and (ii) an elementary step consists of applying  the TM twice, in order to restore periodicity.

The introduction of SO couplings along the bonds, 
see Eqs.~(\ref{eq:hdef}),~(\ref{eq:vdef}), means that the $a_{ik}$ are now
spinors, written on the basis of the eigenvectors of $\sigma^z$ as:
\begin{equation}
a_{ik} = \begin{pmatrix}{a_{ik}^{\uparrow}}\cr{a_{ik}^{\downarrow}}\end{pmatrix}\ ,
\label{eq:spinor}
\end{equation}
where the $a_{ik}^{\uparrow}$, $ a_{ik}^{\downarrow}$ are complex. The matrices
$P_k$ and the subdiagonal $I$ of Eq.~(\ref{eq:matrix}) are now $2N \times 2N$, while the diagonal terms of $P_k$ are doubly degenerate, and the non-zero off-diagonal ones are replaced 
either by the (bond-dependent) negative of matrix $V_{ij}$ of Eq.~(\ref{eq:vdef}) or that of its hermitean adjoint $V_{ij}^\dagger$, 
depending on whether they are supra- or sub-diagonal~\cite{ynik13}.      
The negative unitary  elements of the (now $2N\times 2N$) supradiagonal identity matrix of Eq.~(\ref{eq:matrix}) are replaced by the bond-dependent negative of  $V_{ij}$. 

\section{Metal-Insulator Transition}
\label{sec:mi}

\subsection{Introduction}
\label{sec:mi-intro}

To make contact with previous work on the square lattice~\cite{ez87,e95,aso04}, 
we initially considered systems without site randomness, i.e.,
all $\varepsilon_i=0$, and two versions of SO coupling:
(i) uniform with $\mu=\mu_0 \neq 0$,
$V^x=V^y=V^z=1$ in Eq.~(\ref{eq:hdef}), (ii) random, with  $\mu=2$ 
(as in Ref.~\onlinecite{e95})
and $V^x$, $V^y$, $V^z$ uniformly distributed in $[-1/2,1/2]$.

In case (i), analysis of the resulting dispersion relation shows
that the allowed states occupy a band with the same structure
as that for a system with $\mu \equiv 0$ but, analogously to
the square-lattice systems with uniform SO term of 
Ref.~\onlinecite{e95}, in the 
range $\pm 3\sqrt{1+\mu^2\left[(V^x)^2+(V^y)^2+(V^z)^2\right]}$.

For case (ii) the corresponding density of electronic states 
(DOS) $\rho(\varepsilon)$ can be  evaluated by 
making use of eigenvalue-counting theorems~\cite{dean,dm60,tho72}. 
Our implementation takes advantage of the sparse nature of the
Hamiltonian matrix written on the site basis, and resorts        
to Gaussian elimination algorithms on strip-like          
geometries (with periodic boundary conditions across), 
closely following  the
steps described in Refs.~\onlinecite{sb83,sne86,dqrbs06}. 

Since $\rho(\varepsilon)$ is calculated from the finite difference
between successive values of integrated DOS up to adjacent energies
$\varepsilon$ and $\varepsilon+\Delta$~\cite{sb83,sne86,dqrbs06}, the 
bin size $\Delta$ has to be optimized in order to reduce oscillations 
in the numerical results while still capturing relevant structural 
details of the DOS.
We generally took strip-like systems with $N \ge 80$ sites across,
and length $M \ge 200$, for which $\Delta=0.06$ proved to be a reasonable
choice. For cases such as (ii) where quenched randomness can play a role
in inducing further fluctuations, we saw that for the system sizes
used, the self-averaging provided by having a large(ish) number of
local disorder realizations was enough to render such effects relatively 
unimportant.  

\begin{figure}
{\centering \resizebox*{3.3in}{!}{\includegraphics*{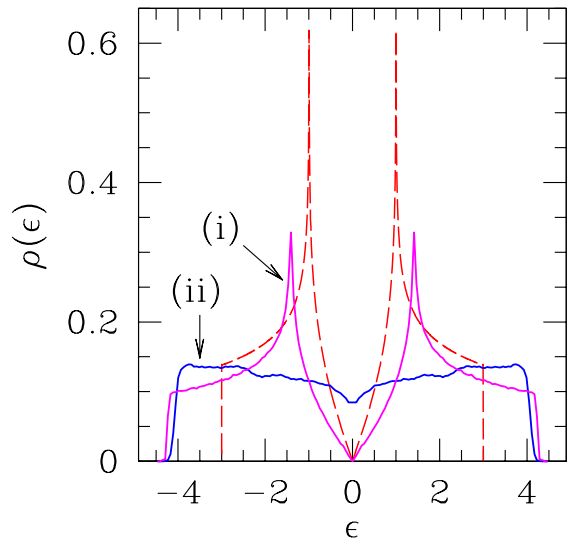}}}
\caption{(Color online) 
Density of states (DOS) for the Hamiltonian Eq.~(\ref{eq:hdef}).
Dashed line, red: $\mu \equiv 0$, exact result. Curves (i) and (ii)
evaluated by eigenvalue-counting on $300 \times 300$ systems.
Curve (i), magenta: uniform SO coupling, $\mu_0=1/\sqrt{3}$.
Curve (ii), blue: random SO coupling, $\mu=2$ (see text).
} 
\label{fig:dos}
\end{figure}

We numerically evaluated $\rho(\varepsilon)$, for 
both (i) [$\,$as an independent check of the soundness of our 
algorithms$\,$] and (ii). Results are shown in Fig.~\ref{fig:dos}.
For (i) we used $\mu_0=1/\sqrt{3}$, which 
ensures that $\mu_0^2\left[(V^x)^2+(V^y)^2+(V^z)^2\right]$ equals
$\mu^2\left[\langle (V^x)^2+(V^y)^2+ (V^z)^2\rangle\right]$ of case
(ii), where angular brackets stand for ensemble average.

One sees that the resulting bands  indeed have
the same width, although the shape of tails at the edges differs. As expected,
these are rather abrupt in (i) and smoother in (ii), in line with the probabilistic
character of the latter's coupling distribution. 
Furthermore, while the band in case (i) keeps
all qualitative features of the $\mu\equiv 0$ system, including the
zero at $\varepsilon=0$, these are lost in case (ii); for example,
the shallow minimum at the origin corresponds to $\rho(0)=0.077(2)$.

Note that the effective strength of intrinsic SO
coupling in graphene is estimated to be $25-50$ $\mu$eV~\cite{kgf10},
of order $10^{-5}$ of the nearest-neighbor hopping 
$\gamma_0=2.8$ eV~\cite{RMP}.
The values used in this Section are for illustration only, and 
not intended to reflect the actual properties of pure graphene.
We return to this point in Sec.~\ref{sec:conc} below.

\subsection{Phenomenological renormalization}
\label{sec:mi-prg}

Here the SO coupling is represented as in model (ii) of 
Sec.~\ref{sec:mi-intro}, with $\mu=2$ (fixed). We now introduce 
site randomness, i.e. $P(\varepsilon_i) = {\rm constant,}\ 
|\varepsilon_i| \leq W/2$ and zero otherwise, and allow the respective 
distribution width $W$ to vary.

Following standard procedures~\cite{ps81} we considered $E=0$,
at the Dirac point of the unperturbed Hamiltonian, and 
iterated the TM on strips of width $N$ sites and length $M \gg N$ with
periodic boundary  conditions across. 
The characteristic Lyapunov exponents were extracted~\cite{ose68,ranmat},
with the longest localization length $\lambda_N$ being given by the inverse
of the smallest of those. According to finite-size scaling  and the 
phenomenological renormalization {\it ansatz}, we plotted
the scaled localization lengths $\Lambda_N \equiv \lambda_N/N$ against
varying $W$, looking for the mutual intersection
of the curves corresponding to different strip widths which gives the 
location of the critical point of the metal-insulator transition.

\begin{figure}
{\centering \resizebox*{3.3in}{!}{\includegraphics*{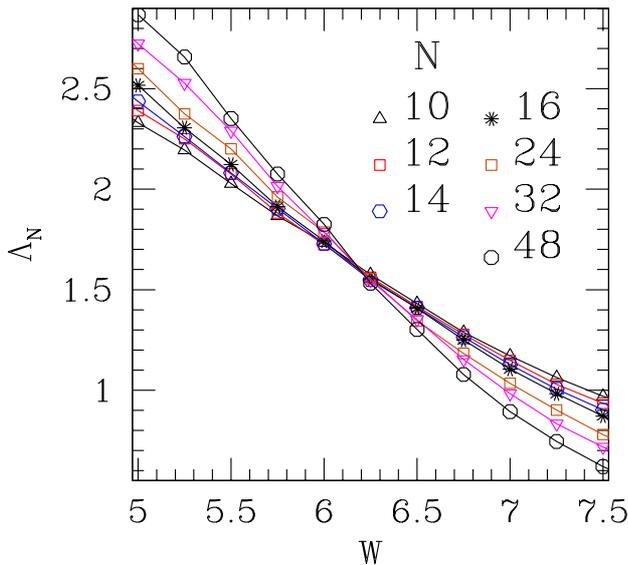}}}
\caption{(Color online) 
Systems with random SO coupling, $\mu=2$: raw data for scaled 
localization lengths against width of site-disorder distribution.
Uncertainties are of order of symbol sizes or smaller.
} 
\label{fig:raw}
\end{figure}

We took $N=10$, 12, 14, 16, 24, 32, and 48, with $M=10^6$. The results
are shown in Fig.~\ref{fig:raw}.

It is known that  nonlinearity of scaling fields and/or irrelevant 
variables~\cite{cardy} can cause sizable distortions in the estimates
of critical quantities at Anderson localization transitions. 
Efficient procedures have been devised to correct for
effects of this sort ~\cite{aso04,so99,sok00}. 
We noted that if only strip widths $N \geq 10$ were used,  
corrections to scaling due to irrelevant fields (usually  dealt with by  methods 
explained in Refs.~\onlinecite{so99,sok00}) were of little import.
Thus we concentrated on accounting for nonlinearities~\cite{aso04}.

Following the lines of Ref.~\onlinecite{aso04} [$\,$see especially their 
Eqs.~(6)--(11)$\,$], we define $W_c$ as the disorder distribution width 
at criticality, and with $w \equiv (W-W_c)/W_c$, we assume that:

\begin{equation}
\ln \Lambda_N =F (N^{1/\nu}\,\psi)\ ,
\label{eq:loglambda}
\end{equation}
where $\psi$ is a smooth function of $E$ and $W$ which goes to
zero at criticality. At fixed $E$, one then considers a truncated Taylor
expansion of  $\psi$ in terms of $w$:
\begin{equation}
\psi= \sum_{k=1}^{n_\psi} \psi_k\,w^k\ .
\label{eq:taylor}
\end{equation}
Nonlinearities in the argument of $F$ are thus taken into account with $n_\psi >1$. 
Plugging this back into Eq.~(\ref{eq:loglambda}) one gets another truncated Taylor series
near criticality:
\begin{equation}
F(x) =\ln \Lambda_c + \sum_{m=1}^{m_0} a_m x^m\  ,
\label{eq:f(x)}
\end{equation}
where $\Lambda_c \equiv \Lambda_{N\to\infty}(E,w=0)$.  One can  either set $\psi_1=1$
in Eq.~(\ref{eq:taylor}), or $a_1=1$ in Eq.~(\ref{eq:f(x)}) without loss of generality. It is expected that by using the logarithm 
in Eq.~(\ref{eq:loglambda}),  the smoothness assumption underlying the Taylor expansions will be fulfilled with fewer terms than if the $\Lambda_N$ themselves were considered.

\begin{figure}
{\centering \resizebox*{3.3in}{!}{\includegraphics*{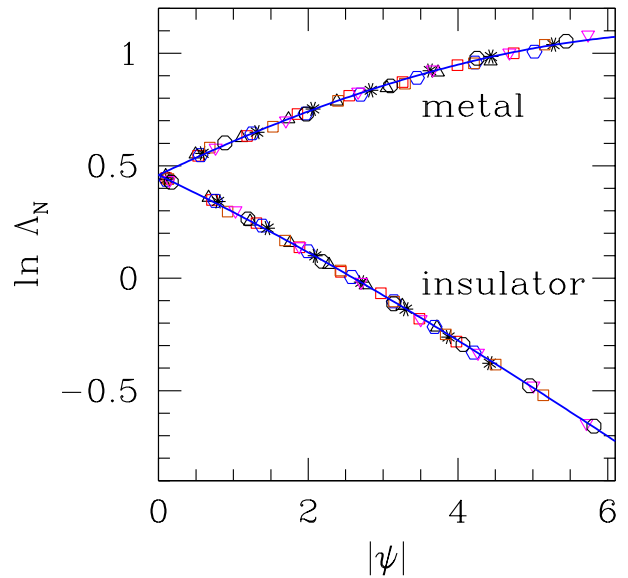}}}
\caption{(Color online) 
Systems with random SO coupling, $\mu=2$: scaling plot of (logarithm of) scaled 
localization  lengths against  $|\psi|$ [$\,$see Eqs.~(\ref{eq:loglambda})--(\ref{eq:f(x)})$\,$].
Key to symbols is the same as in Fig.~\ref{fig:raw}.
} 
\label{fig:scale}
\end{figure}

Fig.~\ref{fig:scale} shows the best-fitting scaling plot of the data of Fig.~\ref{fig:raw},  for which we took $n_\psi=2$, $m_0=3$. Numerical values are given in Table~\ref{t1}. 

\begin{table}
\caption{\label{t1}
Adjusted values of parameters for the scaling plot of Fig.~\ref{fig:scale}, 
with $\psi_1 \equiv 1$ (fixed) in Eq.~(\ref{eq:taylor}).
}
\vskip 0.2cm
\begin{ruledtabular}
\begin{tabular}{@{}ccc}
$\ \ \ \nu$ &\ & \ $\, 2.71 (8)$\ \  \\
$\ \ \ \ln \Lambda_c$ &\ &  $0.459 (9) $\ \   \\
$\ \ \ W_c$ &\ & $6.21(2)$\ \   \\
$\ \ \ \psi_2$ &\ & $ -0.06(1)$\ \   \\
$\ \ \ a_1$ &\ & $-0.158(5)$\ \  \\
$\ \ \ a_2$ &\ & $-0.0076(12)$\ \   \\
$\ \ \ a_3$ &\ & $0.00028(8)$\ \   \\
\end{tabular}
\end{ruledtabular}
\end{table}

The adjusted $\nu=2.71(8)$ is in very good agreement with $\nu=2.75(4)$ of 
Ref.~\onlinecite{aso04} which is, to our knowledge, the most accurate 
result to date . Earlier work gave less accurate estimates, mostly in 
the range $2.6-2.9$ [$\,$see Table 1 of Ref.~\onlinecite{aso02}$\,$].

For comparison of $\ln \Lambda_c=0.459(9)$ with results pertaining to 
the square lattice, one must recall the geometric correction factors of 
the honeycomb lattice~\cite{pf84,bww90,dq06,dq13}.  This means that, in 
order to produce an estimate of the decay-of-correlations exponent 
$\eta= (\pi\Lambda_c)^{-1}$ given by conformal invariance~\cite{cardy}, 
the raw TM result for $\Lambda_c$ must be multiplied by a 
factor of $2/\sqrt{3}$.  We thus obtain $\Lambda_c^{\rm 
corrected}=1.826(15)$ [$\,$so $\eta=0.174(2)\,$] which compares rather well with 
$\Lambda_c=1.844(4)$ of Ref.~\onlinecite{aso04}. We note that the 
critical-amplitude results of 
Ref.~\onlinecite{obuse10} do not compare directly with ours
because those authors study the  scaling behavior of the {\em typical}
localization length.  As is known, this
quantity is given by the zeroth moment of the correlation-function 
probability distribution~\cite{ludwig}, whereas here we deal with {\it average} 
quantities, i.e. ones related to the corresponding first-order moment.

\section{Spin relaxation} 
\label{sec:relax}

In studies of spintronics in semiconductors~\cite{als02,zfds04}
the spin coherence length is one of the quantities of interest.
Here we consider the decay of spin polarization in electronic
transport in quasi one-dimensional geometries, in the presence
of SO couplings. The subject is usually approached via
Green's function techniques~\cite{kka08}. 
We show that this problem can be investigated by 
considering the evolution of wavefunction amplitudes in 
the TM context~\cite{cz97,dq02}.

Usual TM treatments focus on extracting the spectrum of characteristic Lyapunov
exponents,  which demands repeated iteration along $M \gg 1$ columns, with
frequent orthogonalization to avoid contamination~\cite{ps81,ose68,ranmat}.
However, one sees that the recursion relation synthesized in Eq.~(\ref{eq:matrix})
contains information on how the electron wavefunction evolves, starting from
specified initial conditions. In presence of SO couplings the off-diagonal hopping
matrix elements induce spin flips, thus affecting  spin polarization 
along the system's length.

Assume that a fully spin-polarized electron beam is injected into the system, i.e.,
$a_{i0}^{\uparrow},\ a_{i1}^{\uparrow}   \neq 0$,   $a_{i0}^{\downarrow}=a_{i1}^{\downarrow}= 0$, $i=1, \dots, N$.  One can extract  information about the beam's polarization state  $M$ lattice
spacings down the strip by iterating Eq.~(\ref{eq:matrix}) $M$ times and examining the
resulting coefficients $\{a_{iM}^{\uparrow}\}$, $\{a_{iM}^{\downarrow}\}$. In this case
the  beam polarization $\cal P$ at column $M$ is given by:
\begin{equation}
{\cal P}(M)= \frac{\sum_{i=1}^N\{ | a_{iM}^\uparrow|^2-|a_{iM}^\downarrow|^2 \}}{\sum_{i=1}^N \{ | a_{iM}^\uparrow|^2
+ |a_{iM}^\downarrow|^2\}}\  .
\label{eq:pol}
\end{equation}

The initial conditions just mentioned can be viewed as representing an ideal lead, i.e., one
without SO interaction, from which the beam is injected into the strip.  Although,  for
a complete description one would need to take into account reflection at the injecting 
boundary (as well as at the strip's end, presumably linked to a second ideal lead),  these
features do not influence the calculated polarization decay length, as this quantity is computed
from ratios of (sums of squared) amplitudes each taken at a fixed position~\cite{kka08}. 

We evaluate spin polarization profiles on systems with periodic boundary conditions across,
i.e., topologically equivalent to  CNTs; we keep $E=0$, fixed, as this corresponds to 
the Fermi energy which is the relevant level for  transport phenomena. The SO couplings
are   again represented by model (ii) of Sec.~\ref{sec:mi-intro}, although
the overall amplitude $\mu$ will be allowed to vary. Our results
are averages over typically $10^5$ independent samples. For each of these we 
generate random sets of the $\{a_{i0}^{\uparrow}\}$, $\{ a_{i1}^{\uparrow}\}$, as well
as the $\{V_{ij}\}$. Thus we are sampling over the ensemble of steady-flow configurations.

We consider the simplest case  with no site randomness, i.e.,  all $\varepsilon_i=0$ in
Eq.~(\ref{eq:hdef}). This removes one (finite) length scale from the problem, as
one would have the [$\,$site$\,$] disorder-associated mean free path $\Lambda_f \to \infty$. 

\begin{figure}
{\centering \resizebox*{3.3in}{!}{\includegraphics*{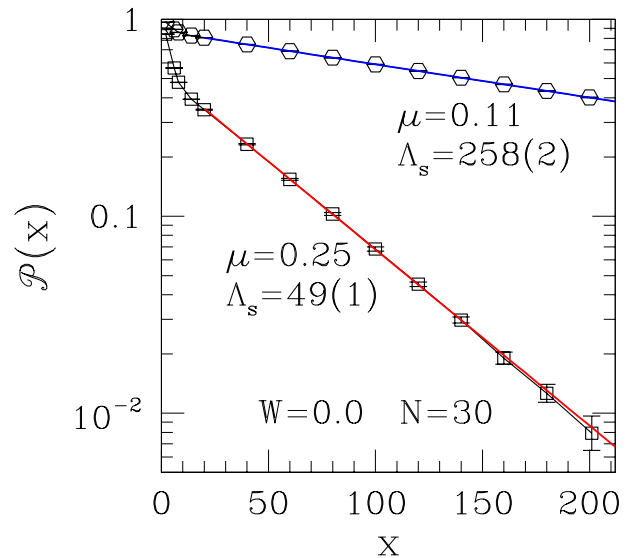}}}
\caption{(Color online)  Log-linear plot of spin polarization ${\cal P}(x)$ versus
position $x$ along axis of nanotube with $N=30$ sites across. A fully-polarized
beam is injected at $x=0$ with $E=0$. Systems with random SO coupling strength 
 $\mu$ as shown,
site-disorder width $W=0$. The characteristic decay lengths $\Lambda_s$ are adjusted values
from fits of data for $x \gtrsim 30$ (fitted lines are shown in color).  
} 
\label{fig:polmu025}
\end{figure}

Fig.~\ref{fig:polmu025} shows that spin polarization ${\cal P}(x)$ settles into exponential decay against position $x$ along the nanotube's axis, after a short transient region of steeper variation.
The characteristic decay lengths $\Lambda_s$ are found by adjusting the appropriate sections of numerical data to a pure exponential-decay form.

\begin{figure}
{\centering \resizebox*{3.3in}{!}{\includegraphics*{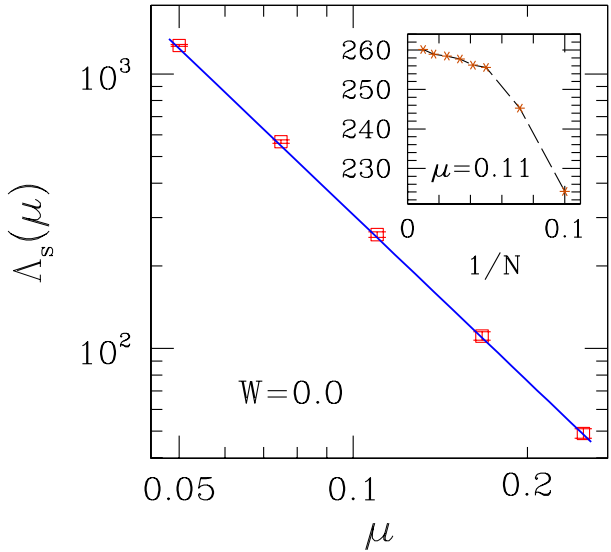}}}
\caption{(Color online)  Main diagram: double-logarithmic plot of
$\Lambda_s(\mu) \equiv \lim_{N \to \infty} \Lambda_s(\mu, N)$ against
 random SO coupling strength  $\mu$, for site-disorder width $W=0$
 and energy $E=0$. Full line is $\Lambda_s(\mu)=3.0\,\mu^{-2.01}$ (best fit to data).
 Inset:  $ \Lambda_s(\mu, N)$ against $1/N$ for $\mu=0.11$.
} 
\label{fig:lambda_s}
\end{figure}

For fixed $\mu$, we found a slight dependence of $\Lambda_s$ on $N$, which is strongest
for small $N \lesssim 20$. As $N$ increases, saturation becomes evident and it is possible
to estimate $\Lambda_s(\mu) \equiv \lim_{N \to \infty} \Lambda_s(\mu, N)$ with good
accuracy by using data for $N$ up to $100$ (see the inset to Fig.~\ref{fig:lambda_s}).

The main diagram in Fig.~\ref{fig:lambda_s} shows that, to a very good extent, the relationship
$\Lambda_s(\mu) \propto \mu^{-2}$ holds for the range $0.05 \leq \mu \leq 0.25$. This is in
line with the findings of Ref.~\onlinecite{kka08} for quantum wires with nonzero on-site disorder
and Rashba-like SO coupling. Such inverse-square dependence  of $\Lambda_s$ on $\mu$ 
is thus likely to be a universal
property for electronic transport in two-dimensional systems with SO effects.

Next, we introduce quenched  impurities with SO interaction onto an otherwise pure
system, i.e. one in which such interaction is generally absent. This amounts to a simplified
representation of pure graphene (where, as recalled in Sec.~\ref{sec:mi-intro},  SO coupling is very weak) doped with suitable impurities. 

Many experimental  and theoretical studies deal with impurities on CNRs,  where edge effects
play an important role in the energetics of favored defect locations .  By considering only
nanotube geometries, here we need not account for this sort of  positional preference 
inhomogeneity. Furthermore,
we restrict ourselves to so-called {\it hole} defects, i.e. adatoms which sit over
the center of a hexagon~\cite{rigo09}.

Denoting by $\rho$ the fraction of randomly chosen hexagons which have an impurity atop 
their center,  we assume that hopping along all six bonds which make such a hexagon is
 characterized by  SO  couplings   $V_{ij}$ as given in Eq.~(\ref{eq:vdef}),
  with the $V^x$, $V^y$,  $V^z$
 randomly distributed as in case (ii) of Sec.~\ref{sec:mi-intro}. We take $\mu=2$ as in
 Sec.~\ref{sec:mi},  which makes the adatom's average SO interaction about as strong
 as the hopping term of pristine graphene.
 
 We consider only $\rho \leq 0.01$ and neglect effects due
 to adjacent impurities, which should be a reasonable approximation in this concentration range.
 
\begin{figure}
{\centering \resizebox*{3.3in}{!}{\includegraphics*{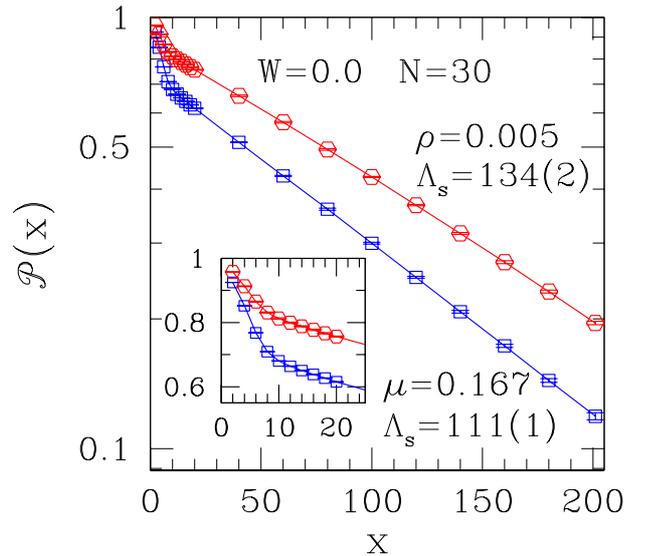}}}
\caption{(Color online)  Main diagram: log-linear plot of spin polarization 
${\cal P}(x)$ versus position $x$ along axis of nanotube  with $N=30$ sites across.
Hexagons (red): SO coupling $\mu=2$ on a fraction $\rho=0.005$ of hexagons,
zero elsewhere (see text); squares (blue): SO couplng $\mu=0.167$ on all
bonds.  Values of $\Lambda_s$ from fits of $x \geq 40$ data to exponential form.
Inset: same data, linear scale on vertical axis, close-up of $x \lesssim 20$ region. 
} 
\label{fig:poluvsr}
\end{figure}

 In Fig.~\ref{fig:poluvsr}
 it can be seen that for $\rho=0.005$ the relaxation length $\Lambda_s$ is very close to that for
 a system with SO coupling on all bonds, and $\mu=0.167$.  For relatively short distances
 $x \lesssim 20$ from the injection edge the initial decay rate is found to be distinctly higher for the latter case than for the former, as highlighted in the inset of Fig.~\ref{fig:poluvsr}. We refrain from ascribing much significance to this difference since we do not expect
 our approach to give an accurate description of such short-range
 effects. So, concerning the region farther than some $20$ lattice spacings from the origin, the effect on (asymptotic) spin polarization decay 
 of $3\%$ of bonds with SO coupling $\mu=2$ is similar to that of all bonds having $\mu=0.167$.

 We checked the dependence of $\Lambda_s$ on $\rho$ and $\mu$.
 For fixed $\mu=2$ we took $\rho=0.003$, $0.005$, $0.0075$, and $0.01$
 which gave a rather good fit to a $\rho^{-x}$  dependence with $x=1.1(1)$; then 
 for fixed $\rho=0.0075$ we additionally made $\mu=0.75$, $1.0$, and $1.5$.
 In this case one gets $\Lambda_s \propto \mu^{-y}$ with $y \approx 1.7$. 
 If one instead assumes that $y=2$, found  for smaller $\mu \lesssim 0.25$ and systems with
 SO couplings on all bonds (see Fig.~\ref{fig:lambda_s}), still  applies here,
 one gets:
  \begin{equation}
\Lambda_s(\rho,\mu) = \frac{A}{\rho\mu^2}\ ,\qquad A=2.4(5)\ ,
\label{eq:rhomu}
\end{equation}
where  the large scatter in the estimate of $A$ reflects the aforementioned poor quality of
the fits of behavior against varying $\mu$. Nevertheless, Eq.~(\ref{eq:rhomu}) can be
a rough-and-ready guide to estimate the relaxation length for values of $\rho$ and $\mu$
closer to physically realizable experiments than those used here.
   
\section{Discussion and Conclusions} 
\label{sec:conc}

We have studied model tight-binding Hamiltonians for the description of electron 
transport in graphene-like geometries, in the presence of SO couplings.  Our main
interest is in the  behavior of very large sheet-like samples, although in the 
quasi-one dimensional context of the TM methods applied here, the
use of periodic boundary conditions across makes our systems topologically akin
to CNTs. For the latter type of system, especially in the limit of very narrow nanotubes,
a more realistic description should probably incorporate curvature effects, 
as done in Ref.~\onlinecite{ando00}.

The energy unit used here equals the nearest-neighbor hopping parameter in pristine graphene,  thus translating into $2.8$ eV~\cite{RMP}. The intrinsic SO coupling in graphene is
estimated as $25-50$ $\mu$eV~\cite{kgf10}; weakly-hydrogenated samples have been reported as giving a colossal enhancement on this by three orders of magnitude~\cite{nat13}. The latter value
would then correspond to $\mu \approx 10^{-2}$,  similar to the low end of the range investigated
in Sec.~\ref{sec:relax}, for systems with SO couplings on all bonds. On the other hand,  we used $\mu=2$, both in Sec.~\ref{sec:mi} and for impurity adatoms in Sec.~\ref{sec:relax},  partly for comparison with extant work~\cite{ez87,e95,aso04}, and also in order to produce well-defined numerical results amenable  to unequivocal interpretation.

In Sec.~\ref{sec:mi} we showed that for strong SO interactions a second-order metal-insulator transition takes place, which is in the same universality class as that found for square-lattice systems~\cite{ando89,ez87,e95,aso04}. Note that the energy $E=0$ used in our calculations
corresponds to the Dirac point of pure graphene, at which (contrary to the square-lattice case) the tight-binding DOS vanishes identically. Although in real graphene a (quantized) non-zero minimum conductivity  has been found at the Dirac point~\cite{mincond}, the effect seen in our results
is unrelated to this, being of a larger order of magnitude (the metallic phase extends up to  site disorder of strength  $W_c \approx 6.2$). In fact, as can be seen in Figure~\ref{fig:dos},  the random SO couplings used in the model investigated in Sec.~\ref{sec:mi-prg}  account for the significant departure from zero  of the DOS close to $E=0$. So, it is the latter feature which sets the stage for the relative robustness of the conducting phase in our model.
The above remark should apply also to the case of 
structures mentioned in Sec.~\ref{intro}, such as boron nitride~\cite{nov05}
or silicatene~\cite{sand14}, which are wide gap insulators. Indeed, it is 
expected~\cite{ekl74}  that in general the introduction of randomness will be
accompanied by the appearance of states outside the pure-system
bands, in particular within the gap. Whether that will be enough 
to give rise to a conducting phase should depend on quantitative
details of the disordered potential.   

In Sec.~\ref{sec:relax} we showed how one can extract information regarding spin polarization decay
with distance from an injection edge,
from the evolution of wave-function amplitudes in a TM context. We illustrated the pertinent ideas
in the simple context of a nanotube-like geometry, with no on-site disorder, and investigated the dependence of the spin relaxation length $\Lambda_s$ on SO coupling strength $\mu$. For small $\mu$, closer to physically realizable values, we found  the dependence $\Lambda_s \propto \mu^{-2}$ which seems to be a universal relationship for two- (or quasi-one) dimensional systems 
with SO interactions~\cite{kka08}. For SO couplings acting only on impurity sites, randomly distributed with  low concentration $\rho \ll 1$, we  found numerically $\Lambda_s \propto \rho^{-1}$,  in line with elementary probabilistic considerations.

It must be noted that modeling SO coupling via Eq.~(\ref{eq:vdef})  with the $V^k$, $k=x,y,z$
randomly distributed in $[-1/2,1/2]$ gives an effect which is, on average, isotropic in spin space.
So this formulation is not suitable,  e.g., for a realistic discussion of precession effects.

Prospects for further application of the ideas presented in Sec.~\ref{sec:relax} would include 
taking  site disorder into account, as well as studying CNR geometries. For the latter, 
inhomogeneities in local current density in the transverse direction to average flow would be
directly accessible.

\begin{acknowledgments}
The author thanks the Brazilian agencies  
CNPq  (Grant No. 303891/2013-0), and FAPERJ (Grants 
Nos. E-26/102.760/2012, E-26/110.734/2012, and E-26/102.348/2013)
for financial support.
\end{acknowledgments}

\end{document}